\documentclass[%
reprint,
amsmath,amssymb,
aps,
]{revtex4-2}

\usepackage{graphicx}
\usepackage{dcolumn}
\usepackage{bm}
\usepackage{siunitx} 

\newcommand{\authoraff}[2]{#1$^{#2}$} 

\begin{document}

\title{Investigating the Ferroelectric Potential Landscape of 3R-MoS$_2$ through Optical Measurements}
%
%

\author{%
\authoraff{\underline{Jan-Niklas Heidkamp}}{1},
\authoraff{Johannes Schwandt-Krause}{1},
\authoraff{Swarup Deb}{1,2,3},
\authoraff{Kenji Watanabe}{4},
\authoraff{Takashi Taniguchi}{5},
\authoraff{Rico Schwartz}{1},
\authoraff{Tobias Korn}{1,*}
}

\affiliation{$^1$Institute of Physics, University of Rostock, 18059 Rostock, Germany}
\affiliation{$^2$Homi Bhabha National Institute, Mumbai, India}
\affiliation{$^3$Saha Institute of Nuclear Physics, Kolkata, India}
\affiliation{$^4$Research Center for Electronic and Optical Materials, NIMS, 1-1 Namiki, Tsukuba 305-0044, Japan}
\affiliation{$^5$Research Center for Materials Nanoarchitectonics, NIMS, 1-1 Namiki, Tsukuba 305-0044, Japan}
\email{Email: tobias.korn@uni-rostock.de}



\begin{abstract}
In recent years, sliding ferroelectricity has emerged as a topic of significant interest due to its possible application in non-volatile, reconfigurable storage devices. This phenomenon is unique to two-dimensional van der Waals materials, where out-of-plane ferroelectric polarization switching is induced by relative in-plane sliding of adjacent layers. The intrinsic stacking order influences the resulting polarization, creating distinct polarization regions separated by domain walls. These regions and the domain walls can be manipulated using an applied vertical electric field, enabling a switchable system that retains the environmental robustness of van der Waals materials under ambient conditions. This study investigates 3R-MoS$_2$ using various optical measurement techniques at room temperature. The spatially resolved optical measurements reveal apparent signal changes corresponding to different ferroelectric stacking orders and variations in layer count. Our findings demonstrate that fast optical mapping at room temperature is a reliable method for probing ferroelectric potential steps in 3R-stacked MoS$_2$ samples, thereby facilitating the identification of the ferroelectric configuration. This approach does not require a conductive substrate or an electrical contact to the sample, making it more versatile than traditional atomic force probe techniques.
\end{abstract}

\maketitle
\section{Introduction}

\begin{figure*}
    \centering
    \includegraphics[width=0.80\linewidth]{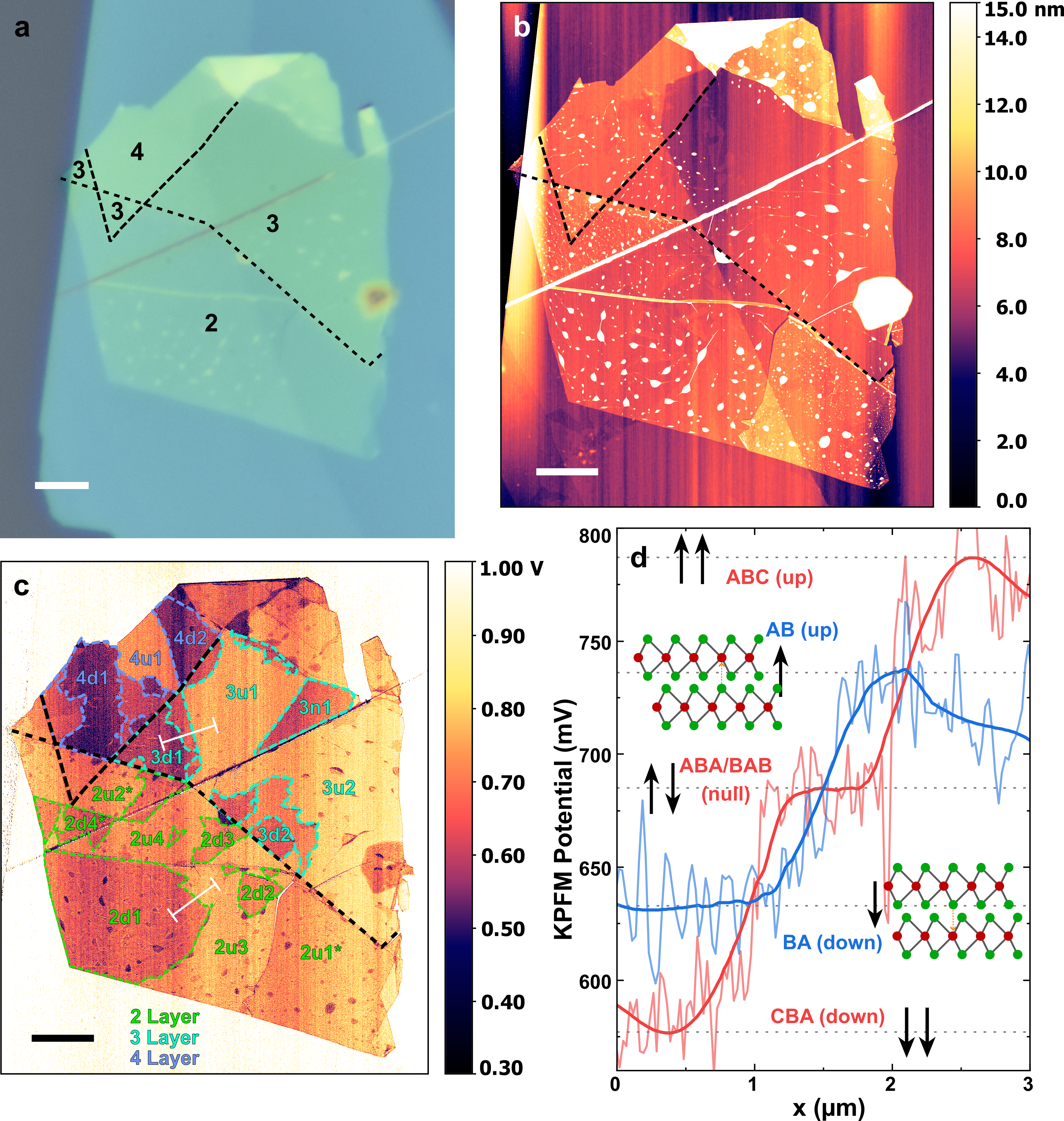}
    \caption{Basic characterization of the 3R-stacked MoS$_2$ sample on top of a hBN backing, the scale bar indicates \SI{5}{\micro m} (a) Optical microscope image with dashed black lines, indicating the borders of the layer regions. The corresponding number marks the layer count of each region. (b) AFM image showing the topography of the flake and the layer outlines (dashed black lines). (c) KPFM map showing the ferroelectric potential landscape across the flake. Regions of equal potential and ferroelectric configuration are identified using specific colors for the layer count, followed by the effective dipole direction (u for up, d for down) and an individual number (e.g., 2d1). The asterisk added to some indicates that these regions have a different optical contrast and amount of adsorbates between hBN and MoS$_2$. (d) Potential profile across a domain edge for a single step in the bilayer region (blue) and a double step in the trilayer region (red). The origins of the profiles are marked in (c) with white lines}
    \label{fig:panel1}
\end{figure*}

The idea of ferroelectric materials is not new, as it was discovered in 1921 by Joseph Valasek in Rochelle salt, which was described as a piezo-electric phenomenon \cite{Valasek1921}.
Naturally, in the last 100 years, the understanding of ferroelectric materials has grown tremendously. They are fundamental in the research towards non-volatile memory, for energy-harvesting and energy-storage devices \cite{Li2017, Youngblood2023}. Tradionally used are three-dimensional (3D) perovskite oxides including BiFeO$_3$ or LiNbO$_3$ and ceramic-like materials like Barium Titanium-Bismuth Zinc (BT-BZNT) \cite{Dawber2005, Salazar2022, Chen2018, Rabson1995, PeiyaoZhao2019}.
The ferroelectric surface potential can also be used in field-effect transistors to increase their on-off ratio \cite{Si2018}. Two-dimensional (2D) semiconducting materials like transition metal dichalcogenides (TMDCs) are ideally suited for such devices, as they mostly consist only of surfaces and are highly tunable by electric gating \citep{Ross2013, Ciarrocchi2019, Lu2017, Salazar2022}.
Similarly, the optical properties of TMDCs can also be influenced by adjacent ferroelectric layers \cite{Soubelet2021}.
A particularly exciting development in ferroelectric materials is the emergence of sliding ferroelectricity \citep{Liu2016, Hu2019, Wu2021, Yasuda2021, Stern2021, Zhou2024}, where the polarization arises not from ion displacement within the unit cell of a 3D crystal but due to inversion asymmetric atomic registry at the interface between van der Waals-coupled layers. This mechanism, enabled by the weak interlayer bonding characteristic of 2D materials, gives rise to polarization states that are robust and reversible.
Among the materials exhibiting this behavior, parallel-stacked hexagonal boron nitride (hBN) has already been studied extensively \citep{Stern2021, Yasuda2021, Fraunie2023}. It is created via tear-and-stack or folding techniques and was investigated regarding its ferroelectricity and response to electrical gating \citep{Stern2021, Yasuda2021}.
The potential landscape introduced by the different ferroelectric polarizations in the parallel-stacked hBN also affects the local charge carrier concentration of neighboring TMDCs like WSe$_2$, facilitating optical measurements to probe the polarization landscape of the material. \cite{Fraunie2023}.
Another ferroelectric 2D material is rhombohedrally stacked (3R-) MoS$_2$, which has garnered significant attention and can be chemically synthesized in single crystals \citep{Ullah2021}. Unlike the more common 2H phase, the 3R-MoS$_2$ exhibits interfacial ferroelectricity, emerging from the rhombohedral parallel stacking of the layers.
With the increase in layer count, the number of possible configurations increases, as each interface can contribute ferroelectric potential steps with different polarity \cite{Deb2022, Weston2022, Meng2022, Deb2024}. 
The discovery of ferroelectric switching via interlayer translation in such systems has introduced a new class of tunable materials for low-power, scalable electronics \citep{Li2017, Youngblood2023, Bian2024}.\\
In this study, we expand on ways to characterize ferroelectric TMDCs. Thus far, ferroelectricity in vdW materials has mostly been studied using variations of scanning probe microscopy, including piezoresponse force microscopy (PFM) and scanning tunnelling microscopy (STM) \cite{Weston2022, Molino2023, Ko2023}. In some studies, optical techniques like second harmonic generation microscopy, or transport measurements have also been used \cite{Zhang2022, Liang2022, Yang2024, Liang2025-Phys, Liang2025-Nat}. However, with the new sliding ferroelectrics, the technique of choice is Kelvin probe force microscopy (KPFM) \cite{Deb2022, Deb2024}, an atomic force microscopy (AFM)-based measurement of the surface potential. TMDCs and all van der Waals materials, in a nutshell, consist mostly of surfaces; therefore, this technique can be perfectly used to investigate the ferroelectric potential landscape of sliding ferroelectrics such as 3R-MoS$_2$. However, these techniques require specific sample geometries, making them impractical in many use cases.
Here, we expand the understanding of ferroelectric MoS$_2$ concerning optical measurement techniques such as micro-photoluminescence spectroscopy and micro-Raman spectroscopy, and map the changes of the optical spectra caused by the changes in ferroelectric potential. The aim is to find ways to characterize sliding ferroelectric samples without the requirement for a specific sample structure, allowing an optical investigation and mapping of the ferroelectric potential landscape.
Remarkably, we find clear correlations between optical spectroscopy signatures and the ferroelectric structure of 3R-MoS$_2$, enabling room-temperature mapping of ferroelectric domains.


\section{Results and Discussion}

The sample investigated here consists of a 3R-MoS$_2$ flake on top of hBN, which was cleaved directly onto a p-doped silicon wafer with a 90~nm layer of SiO$_2$. During preparation, the thickness of the hBN was estimated using its interference color \cite{Anzai2019} and later confirmed by AFM.
The MoS$_2$ flake is exfoliated from a bulk 3R-MoS$_2$ crystal (HQ Graphene) onto polydimethylsiloxane (PDMS) and transferred with a variation of the dry deterministic transfer technique described in \cite{Castellanos2014}.
Hereby, the freshly exfoliated 3R-MoS$_2$ is placed on top of the clean surface of the cleaved hBN, resulting in little PDMS residue between the flakes.
A more detailed description of the sample preparation can be found in the Methods section.\\
The layer count of the MoS$_2$ flake was initially determined by its contrast on the PDMS.
After transfer, the layer count was confirmed with non-contact-mode (NCM) AFM scans as well as with a Raman scan using a $\SI{532}{nm}$ excitation laser. Here, the frequency difference between the E$^1_{2g}$ and the A$_{1g}$ modes changes with the number of layers and, therefore, shows regions of constant thickness. This frequency difference shift is well-established for 2H-stacked MoS$_2$ \cite{Lee2010} and behaves similarly for 3R stacking.\\
\noindent The ferroelectric domain structure in 3R-MoS$_2$ can not be distinguished with a usual optical microscope. To resolve this, commonly KPFM measurements are performed, where the surface potential is probed using an AFM tip and an oscillating electric potential applied between the tip and the measured sample. For reliable results, the tip needs to be in close proximity to the top surface. Therefore, the sample needs to reside either on a conductive substrate, or the insulating layer between the conducting surface and the sample needs to remain thin enough for a good imaging contrast. In our case, a thin oxide–coated p-doped Si wafer was employed to minimize charge transfer arising from contact potential differences between the sample and the substrate. This approach ensures that the observed domain-specific changes reflect the intrinsic properties of the system, rather than artifacts introduced by ferroelectricity-mediated sample–substrate interactions. In addition, no top hBN encapsulation was used, thereby preserving close proximity between the tip and the sample.  
The sample under consideration is relatively large, with three topographically distinct regions visualized in Figure \ref{fig:panel1}(a) with an optical microscope image. The layer regions are outlined with dashed black lines, and the layer count is indicated by the number in each region. The topography of the sample can further be visualized through the AFM image in Figure \ref{fig:panel1}(b), showing the z-height of the flake with the layer regions outlined with dashed black lines.
\noindent The ferroelectric potential landscape is mapped using KPFM to gain a reference for later optical measurements. This mapping is shown in Figure \ref{fig:panel1}(c). Here, the typical potential steps across the ferroelectric domain edges \cite{Deb2022} are visualized in Figure \ref{fig:panel1}(d). These steps in surface potential are indicative of a change in stacking order, therefore, ferroelectric polarization. 
Below, we explain the nomenclature we followed to index the domains observed in the KPFM map.\\
The integer at the beginning indicates the layer count of a given domain. The total intrinsic ferroelectric polarization is denoted as ``u'' (or ``up'') for polarization pointing upward, ``d'' (or ``down'') for polarization pointing downward, and ``n'' (or ``null'') for zero net polarization. Null domains correspond to stacking configurations in which the combined polarizations of individual interfaces cancel, yielding zero total ferroelectric polarization within the stack. This cancellation requires an even number of interfaces, which occurs in regions with an odd number of layers. For example, a bilayer region with a single interface can host only up or down orientations, while a trilayer region with two interfaces can additionally exhibit the null orientation (see Figure \ref{fig:panel1}(d)). Notably, because the up and down states in the trilayer arise from cumulative contributions of two interfaces, they exhibit a stronger ferroelectric potential. A tetralayer region, with yet another interface, allows more possible configurations—for instance, three-up and two-up–one-down. Both correspond to overall up polarization, but with different potential magnitudes. However, in our sample, the tetralayer regions are relatively small, so few variations are observable. Moreover, the increased thickness reduces the signal-to-noise ratio, making it harder to resolve distinct potential steps.\\
As an illustration of our notation, consider the label 3d1. This refers to a downward-polarized domain in a trilayer region, with the final digit as a user-defined index to uniquely identify that domain on the map with the same layer count. 
Our naming scheme differs from the convention commonly used in the literature \cite{Stern2021, Deb2022, Fraunie2023, Deb2024}. Whereas prior works typically emphasize stacking order, we explicitly focus on the orientation of the intrinsic dipole moment. The two approaches, however, can be readily reconciled. For example, in the bilayer region, the “down'' domain corresponds to BA stacking and the “up'' domain to AB stacking. 
Analogous correspondences extend to trilayer and tetralayer regions, accounting for the additional layers and possible stacking sequences.\\
Some domains are additionally marked with an asterisk. These highlight regions of the sample, where the adsorbates between hBN and the 3R-MoS$_2$ flake differ substantially. This can be identified in the AFM image in Figure \ref{fig:panel1}(b), as these regions have a different height, and also in the KPFM image, they have a slight difference in surface potential, despite having the same ferroelectric orientation. \\

\begin{figure}
    \centering
    \includegraphics[width=0.95\linewidth]{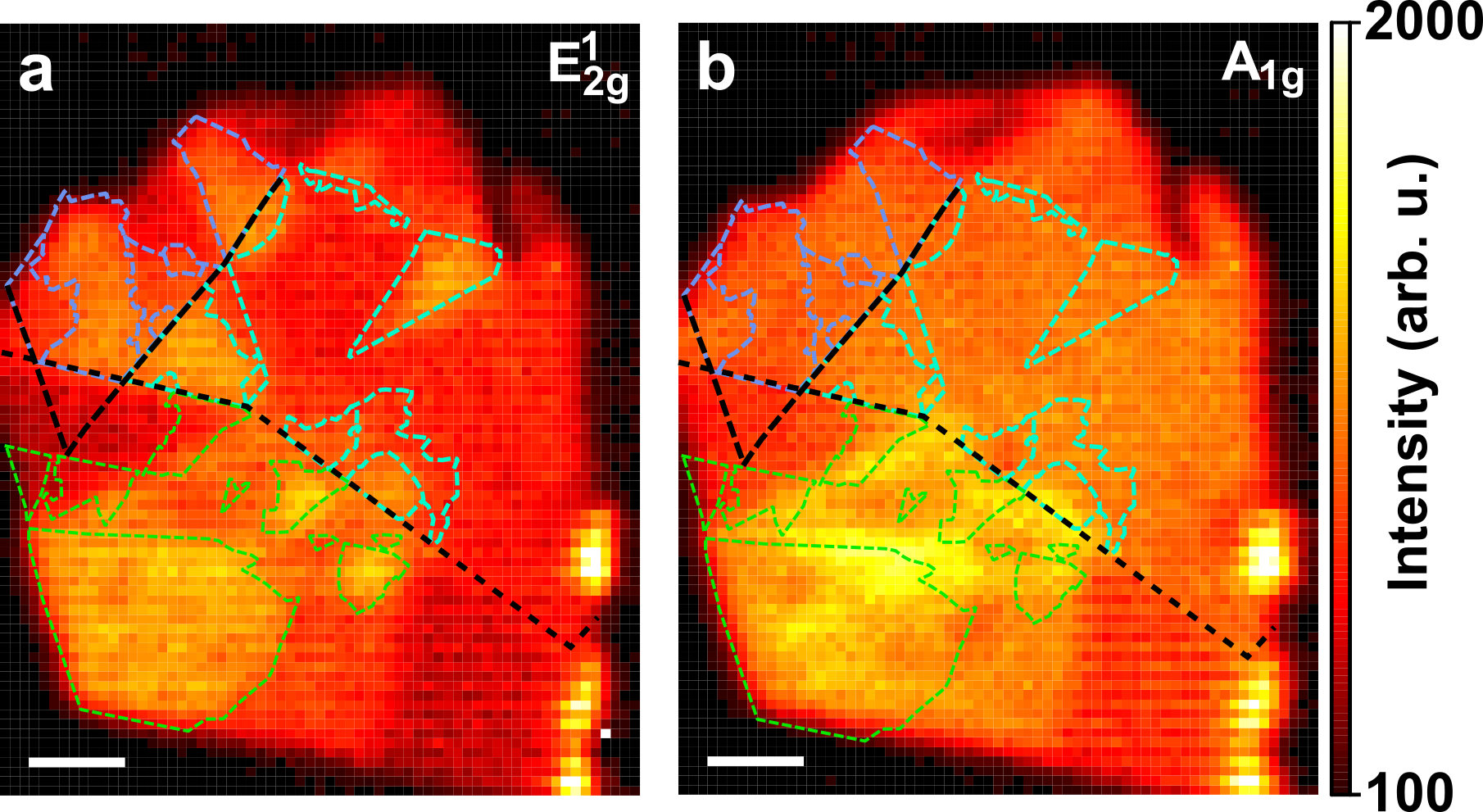}
    \caption{Intensity distributions of (a) the E$^1_{2g}$ Mode (at \SI{383}{cm^{-1}}) and (b) the A$_{1g}$ Mode (at \SI{405,8}{cm^{-1}}) extracted from a \SI{633}{nm} exitation near-resonant Raman scan. The white scale bar indicates \SI{5}{\micro m} and the domain and layer outlines are marked according to the previously introduced scheme (see Figure \ref{fig:panel1}(c)).}
    \label{fig:panel2}
\end{figure}

To investigate the optical responses of the ferroelectric domain landscapes, near-resonant Raman spectroscopy was utilized.
As previously reported for 2H-stacked MoS$_2$ \cite{Livneh2015, Lee2015}, near-resonant excitation yields a more complex Raman signal than off-resonant excitation, with a multitude of second-order Raman modes becoming visible besides the first-order E$^1_{2g}$ and A$_{1g}$ modes that are present for non-resonant excitation.
We note that near-resonant conditions for MoS$_2$ are achieved using excitation at a wavelength of \SI{633}{nm}, which is among the most common wavelengths used in commercial micro-Raman spectroscopy setups due to the prevalence of HeNe lasers. 
Figure \ref{fig:panel2} shows spatially resolved Raman intensity maps of the E$^1_{2g}$ and the A$_{1g}$ modes. The Raman intensity of the E$^1_{2g}$ mode shows a clear correlation to the domain structure indicated by the dashed outlines. By contrast, the A$_{1g}$ appears to be less sensitive to the change of intrinsic ferroelectric potential, and seems not to correlate with the trilayer and tetralayer regions.

\begin{figure}
    \centering
    \includegraphics[width=0.95\linewidth]{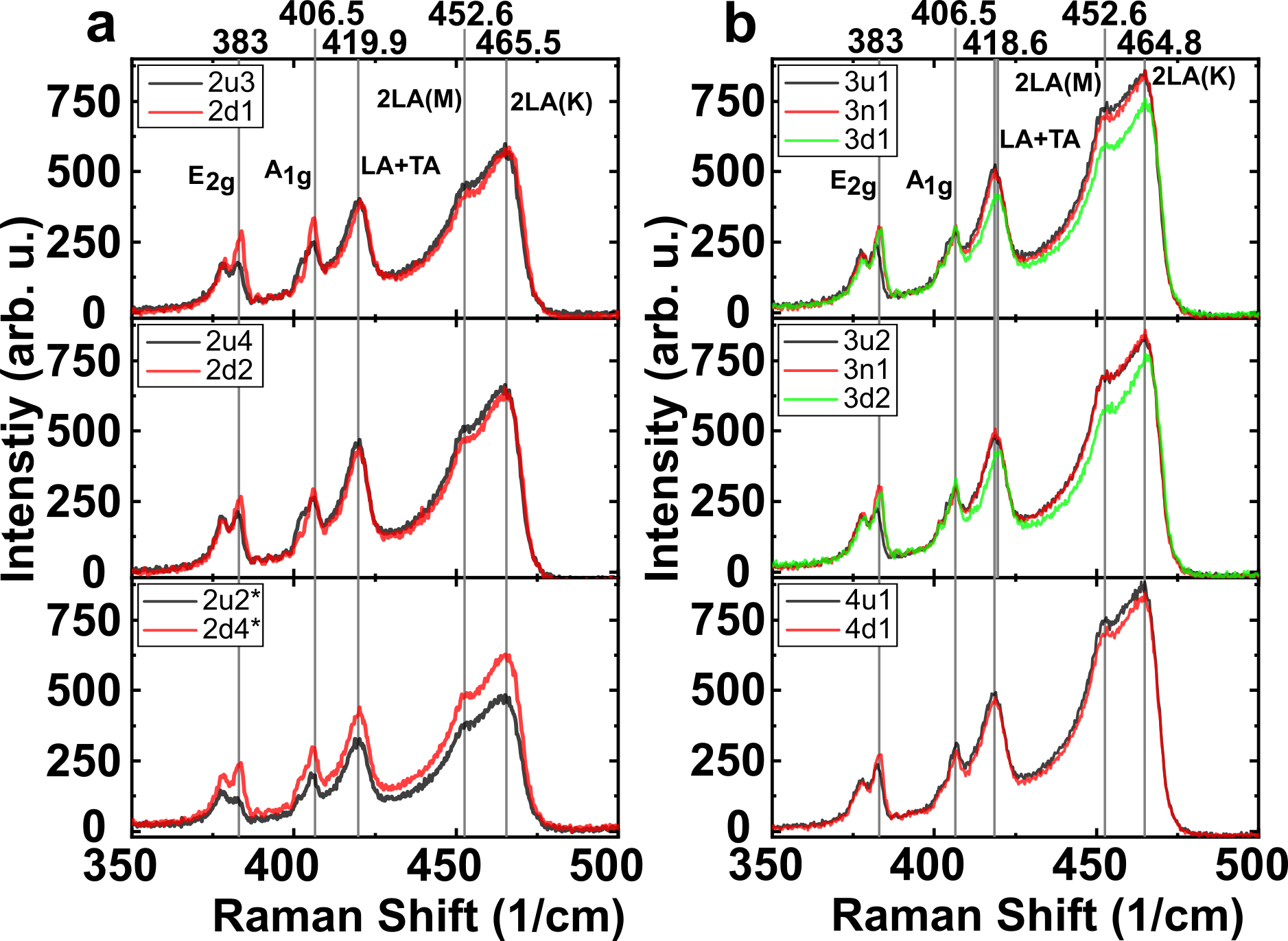}
    \caption{Changes of spatially averaged \SI{633}{nm} excitation Raman signal between neighboring domains of opposing configurations with the same layer count. (a) Comparing opposing domains in the bilayer region. The domains in the bottom frame are marked with an asterisk to indicate that their location on the sample has a different adhesion to the hBN than the rest.  (b) Comparing opposing domains in the trilayer and tetralayer regions. The trilayer spectra include the null domain specific to regions of odd layer count. For the identification of the second-order Raman modes, several papers on near-resonant Raman spectroscopy on 2H-MoS$_2$ were referenced \cite{Lee2015, Livneh2015, Lu2017, Carvalho2017}.}
    \label{fig:panel3}
\end{figure}

\noindent The measured Raman spectra presented in Figure \ref{fig:panel3} are slightly tainted by the superimposed photoluminescence signal of the flake and an etaloning effect occurring within the experimental setup. \\
The presented Raman spectra were improved by removing the background PL and the etaloning effect, as well as spatially averaging inside the domains over a matrix of 5 by 5 measurement spots. For a more detailed description of the data processing steps, see supplementary information Figure \ref{fig:SI_DataProcessing}.
When looking into the near-resonant Raman signal of the sample in Figure \ref{fig:panel3} and comparing neighboring domains with opposing polarizations in the same layer regions, the intensity change of the Raman features becomes apparent. 
As evident from Figure \ref{fig:panel3}(a), in the bilayer region, the E$^1_{2g}$ mode has enhanced intensity in the down domain. The A$_{1g}$ mode exhibits a similar trend, albeit being less pronounced. This holds both in terms of absolute intensities and intensity ratio of the first-order modes to the second-order ones; consider, for example, the ratio of E$^1_{2g}$ to LA+TA or 2LA(K).
In the trilayer and tetralayer regions (see Figure \ref{fig:panel3}(b)), a similar behavior is observed; however, the effect is strongly reduced. Here, the E$^1_{2g}$ mode still shows the increase in the down domain, but this effect is absent in the A$_{1g}$ mode, indicating a layer-dependent difference in the response of the two modes. 
In contrast, the distinction between the trilayer up and the trilayer down domain is evident from the reduced intensity of LA+TA and both 2LA modes, along with the most sensitive E$^1_{2g}$ mode. Interestingly, the LA+TA mode also exhibits a discernible blueshift of \SI{1}{cm^{-1}} in the down domains.
\\

\begin{figure}
    \centering
    \includegraphics[width=0.95\linewidth]{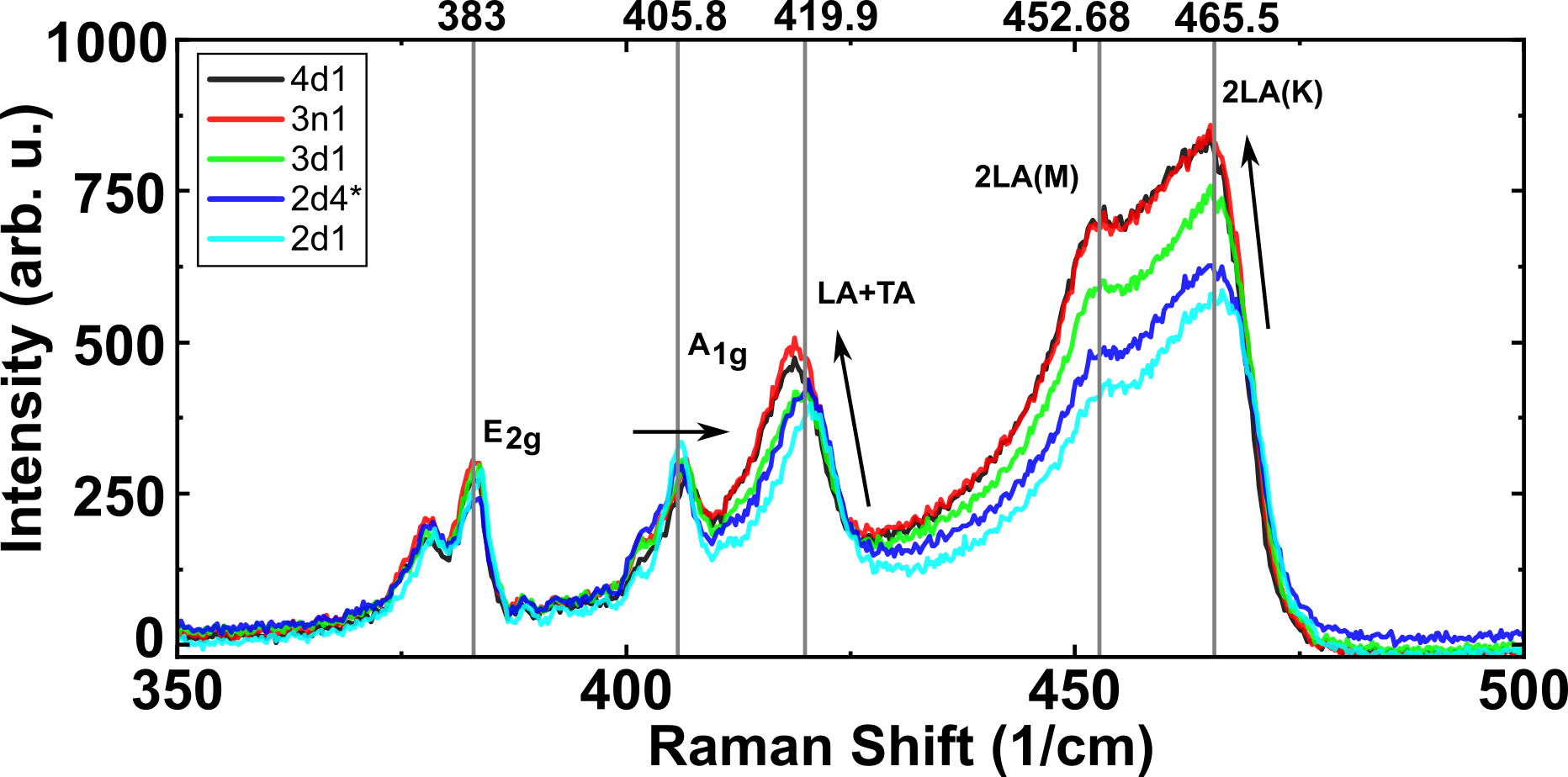}
    \caption{Change of the spatially averaged \SI{633}{nm} excitation Raman signal of the down domains across the bi-, tri-, and tetralayer regions. The spectrum of a trilayer null domain is included for comparison. The arrows visualise the spectral changes with increasing layer count.  Note the increase in intensity of the Raman features around \SI{450}{cm^{-1}} but decrease in the response of the E$^1_{2g}$ and the A$_{1g}$ modes as they do not increase in intensity along the overall signal. The domain marked with an asterisk is located in a region with different adhesion to the hBN than the rest.}
    \label{fig:panel4}
\end{figure}

\noindent We attribute the domain-specific Raman changes to variations in the itinerant charge carrier population and the spectral shift of the absorption, driven by local ferroelectric dipole moments that funnel background electrons across the layers and at the sample-substrate interface \cite{Liang2022, Deb2024, Liang2025-Phys}. However, changes in carrier density alone cannot fully explain our observations, since the asymmetric intensity variation between the E$^1_{2g}$ and A$_{1g}$ peaks deviates from earlier results on externally gated 2H-MoS$_2$ \citep{Lu2017}, where the E$^1_{2g}$ and A$_{1g}$  intensities grow/fall equally with changing gate voltage. This discrepancy suggests that one of these modes couples more strongly to either the stacking order or the intrinsic dipole moment, thereby introducing an additional domain dependence. We also note that the observed behavior is not solely caused by electron-phonon coupling, as off-resonant Raman measurements using 532~nm (2.33~eV) did not show a comparable response to the ferroelectric potential landscape as the near-resonant measurements did. Developing a microscopic theory to capture these effects remains an open challenge.


\begin{figure}
    \centering
    \includegraphics[width=0.95\linewidth]{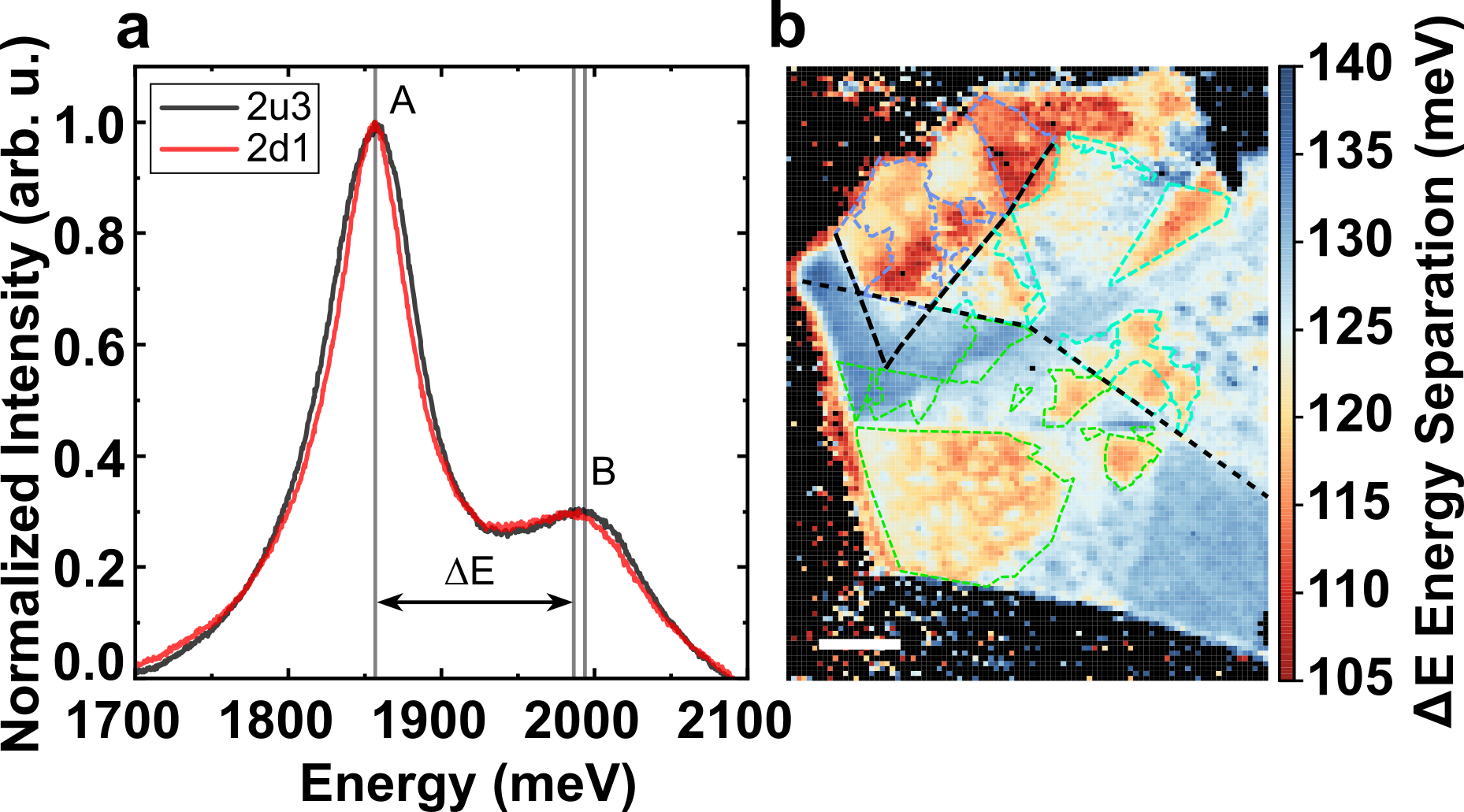}
    \caption{(Room temperature PL with \SI{532}{nm} excitation to investigate the ferroelectricity-influenced behavior of the A and the B excitons (at \SI{1855}{meV} and \SI{1990}{meV}). (a) Comparing spatially averaged and normalized spectra of down and up domains in the bilayer region of the 3R-MoS$_2$ flake. The down domain has a reduced half-width of the A exciton with and the B exciton is redshifted by about \SI{10}{meV}. (b) Spatial map of the energetic distance of the A and B excitons based on fitted data. Peak distance changes with domain orientation across all layer counts in the same fashion. Outlines of the ferroelectric domains are marked in the previously introduced scheme. The white scale bar indicates \SI{5}{\micro m}.}
    \label{fig:panel5}
\end{figure}

\noindent To verify the doping dependencies observed in the near-resonant Raman investigations, photoluminescence (PL) spectroscopy was performed, which is a commonly used spectroscopy tool to estimate charge carrier concentrations. Controlling charge carrier density in gated structures has established this relation \citep{Liang2022}.
Here, a $\SI{532}{nm}$ excitation laser is used for spatially resolved room temperature micro-photoluminescence maps.
Our discussion of PL signals is divided into two parts. First, we focus on the high-energy signals arising from intralayer A and B excitons; then, we will address the low-energy signal, which stems from momentum-indirect and interlayer exciton transitions. 
In Figure \ref{fig:panel5}(a), the photoluminescence signals of the down and up domains of the bilayer region are spatially averaged and normalized. Notably, the reduced half-width of the A exciton as well as the redshift of the B exciton is evident in the down domain.
This redshift is visualized in Figure \ref{fig:panel5}(b) by use of an automatic fitting routine and mapping the energetic distance $\Delta E$ between the A and B excitons for every position. The color scheme indicates the redshift of the B exciton as red regions, which coincide with the down domains of the ferroelectric landscape indicated by the dashed outlines shown first in Figure \ref{fig:panel1}(d).
Comparing the regions in Figure \ref{fig:panel5}(b) to the outlines of the ferroelectric domains, it becomes apparent that the achieved contrast of the ferroelectric domain mapping using this method is far superior to the near-resonant Raman maps.
This redshift of the B exciton in the down domain is consistent with the PL response on externally gated samples \cite{Liang2022}.\\ 

\begin{figure*}
    \centering
    \includegraphics[width=0.95\linewidth]{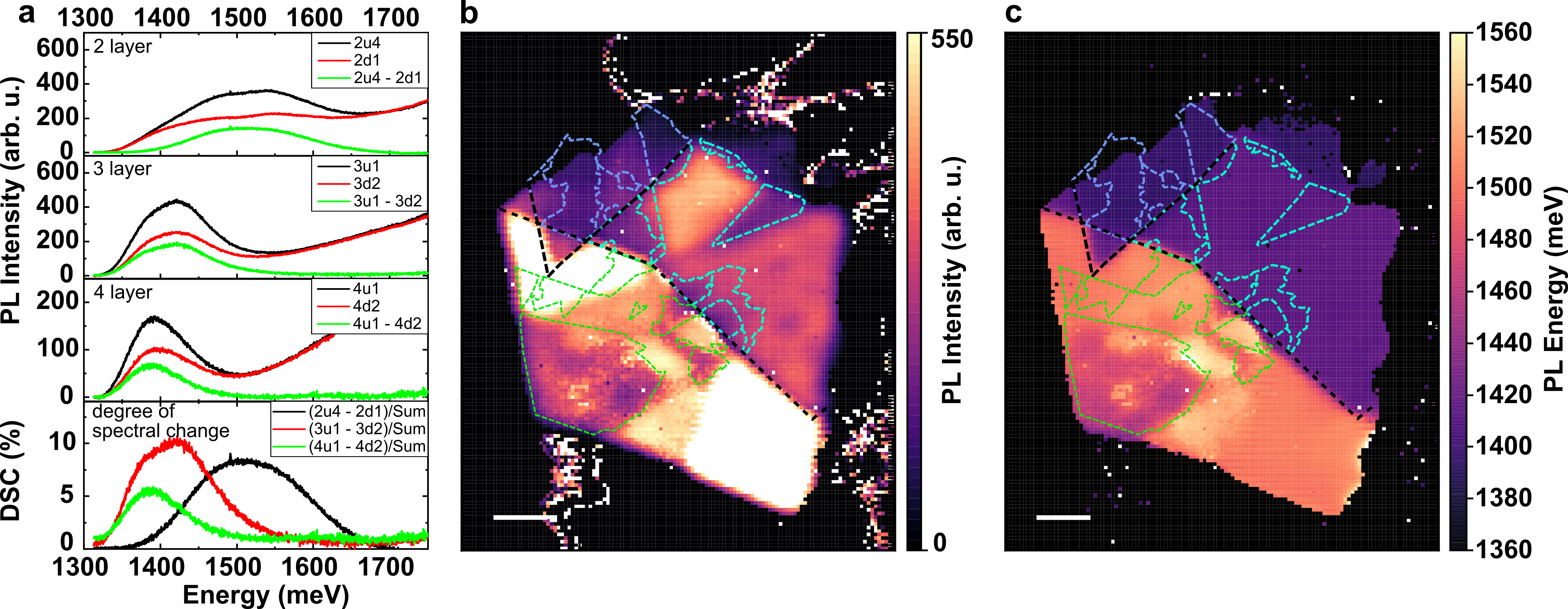}
    \caption{Room temperature PL with \SI{532}{nm} excitation to investigate the low-energy features of multilayer 3R-MoS$_2$ (a) Comparing the spatially averaged photoluminescence signal for opposing domains in bi-, tri-, and tetralayer regions and their spectral change. The bottommost frame compares the degree of spectral change (DSC) on each layer region. (b,c) Maps of the PL intensity and the center energy of the low-energy feature based on an automatic fitting routine. Layer and domain regions are indicated via the previously established scheme, and the scale bars indicate \SI{5}{\micro m}.}
    \label{fig:panel6}
\end{figure*}

Next, we focus on the investigation of the low-energy photoluminescence signal of the 3R-MoS$_2$ and compare its changes with the switch in ferroelectric configuration. 
Before comparing the spectral change across different ferroelectric configurations, it is noted that there is a distinct change of the low-energy feature with a change in layer count (see Figure \ref{fig:panel6}(a)).
The low-energy feature of the bilayer region has an asymmetric double peak centered around 1.5~eV, and with an increase in layer count, the separation of the double peak decreases, the feature redshifts, and overall intensity decreases. These layer-dependent PL intensity and PL energy are visualized in the false-color maps in Figure \ref{fig:panel6} (b,c), where the intensities and center positions are mapped using an automatic fitting routine.
To compare opposing ferroelectric domains inside the same layer region, spatially averaged spectra with similar background are chosen and plotted in Figure \ref{fig:panel6}(a) along with their spectral change (intensity difference of up and down domains). The bottommost frame compares the degree of spectral change (DSC) across the different layer regions. The DSC is calculated by taking the spectral change and dividing it by the sum of both spectra. Multiplying this ratio by 100 yields the percentage of the spectral change resulting from the flip of the ferroelectric configuration. For the trilayer region, the DSC is the highest with around 10~\%. 
The PL intensity map in Figure \ref{fig:panel6}(b) gives insight into the ferroelectric domain structure of all layer regions established with KPFM measurements. With a higher layer count, the apparent contrast in the false-color map is reduced due to the lower overall PL intensity.
The center energy of the low-energy feature is mapped in Figure \ref{fig:panel6}(c), and due to the energy shift with an increase in layer count, the layer regions are visible. However, the ferroelectric domain structure can not be differentiated with just the center energy.

\section{Conclusion}

The goal of this work was to investigate the ferroelectric-potential-related changes in the optical signal of 3R-MoS$_2$ regarding near-resonant Raman and PL spectroscopy. The near-resonant Raman investigations showed a significant intensity change of the first-order E$^1_{2g}$ and the A$_{1g}$ modes, as well as an intensity modulation of the second-order Raman modes in the trilayer regions.
These effects correlate with the spectral shift of the absorption of the 3R-MoS$_2$ caused by the change in charge carrier concentration.
The intensity modulation of the first-order Raman modes can be used to map and identify the ferroelectric potential landscape.  
The room temperature photoluminescence shows a change in the half-width of the A exciton and an energy shift of the B exciton with the domain structure in the bilayer region. The energy shift can be mapped, resulting in a ferroelectric domain mapping with contrast not dissimilar to the KPFM measurements. 
Also, the position of the low-energy PL emission in 3R-MoS$_2$ gives a clear indication of the layer count, distinctly separating the bilayer, trilayer, and tetralayer regions.
The intensity of this emission is modulated by the ferroelectric configuration of the 3R-MoS$_2$ as well, yielding another means of identifying ferroelectric domains via optical spectroscopy.
All measurements shown here can easily be performed at room temperature using commercial off-the-shelf micro-Raman spectroscopy systems and allow mapping ferroelectric domain structures in 3R-MoS$_2$ without needing to resort to KPFM measurements. 

\section{Methods}
\textbf{Exfoliation}\\
First, \SI{90}{nm} SiO$_2$ wafer pieces are prepared and cleaned by baking them for 20 minutes at \SI{360}{\degree C} on a hotplate. Two different types of exfoliation are used. \\
The hBN is exfoliated directly from Scotch tape onto the prepared wafer pieces. Before the Scotch tape with the bulk hBN is removed, the wafer pieces are heated to \SI{110}{\degree C} to increase the adhesion of the hBN to the SiO$_2$. After the exfoliation, the wafer pieces are baked again for 20 minutes at \SI{360}{\degree C} to remove residue from the Scotch tape. \\
The MoS$_2$ few-layer flakes are exfoliated from bulk 3R-MoS$_2$ (HQGraphene) using PDMS. The thickness is determined optically via the relative color contrast with the PDMS background.\\

\textbf{Transfer}\\
The MoS$_2$ flake is transferred onto a selected hBN flake using a variation of the dry deterministic transfer technique described in \cite{Castellanos2014}.\\

\textbf{KPFM/AFM}\\
All KPFM and AFM measurements were performed using the Park Systems NX20 using PointProbe Plus Electrostatic Force Microscopy n-doped tips (PPP-EFM). The sample was first mapped in non-contact mode to record the topography, and second, the area was scanned again to measure the KPFM signal. The two built-in lock-in amplifiers are used to measure the sideband frequencies.\\

\textbf{Spectroscopy}\\
Micro-photoluminescence and micro-Raman measurements were performed in self-built microscopy setups. In both setups, the samples were excited with either a \SI{532}{nm} (\SI{2.33}{eV}) frequency-double solid-state laser or a \SI{633}{nm} (\SI{1.96}{eV}) diode laser focused to a spot size of about \SI{1.5}{\micro m} using a 50x microscope objective. 
For both measurements, a spectrometer (SpectraPro HRS-750) with 750 mm focal length was used; however, the PL measurements used a 150 grooves/mm grating while the Raman measurements used a 1800 grooves/mm grating.
In both cases, a Peltier-cooled CCD camera (Princeton Instruments PIXIS 400) mounted to the spectrometer was used to resolve the spectral data. For mapping, the sample was mounted on a motorized xy stage and scanned beneath the fixed optical beam path.

\section*{Acknowledgements}
The authors gratefully acknowledge technical assistance by I. Barke. 
T.K. acknowledges financial support by the DFG \emph{via} the following grants: SFB1477 (project No. 441234705), KO3612/5-1 (project No. 398761088), and KO3612/8-1 (project No. 549364913). 
K.W. and T.T. acknowledge support from the JSPS KAKENHI (grant numbers 21H05233 and 23H02052) and World Premier International Research Center Initiative (WPI), MEXT, Japan.

\section*{Conflict of Interest}
The authors declare no conflict of interest.

\bibliography{bibliography}

\bibliographystyle{ieeetr}

\clearpage
\onecolumngrid
\newpage
\renewcommand{\thefigure}{S\arabic{figure}}
\renewcommand{\thetable}{S\arabic{table}}
\setcounter{figure}{0}
\setcounter{table}{0}

\section{Supplementary information}
\subsection{Details of Data Processing}

\begin{figure}[h]
	\centering
	\includegraphics[width=0.85\linewidth]{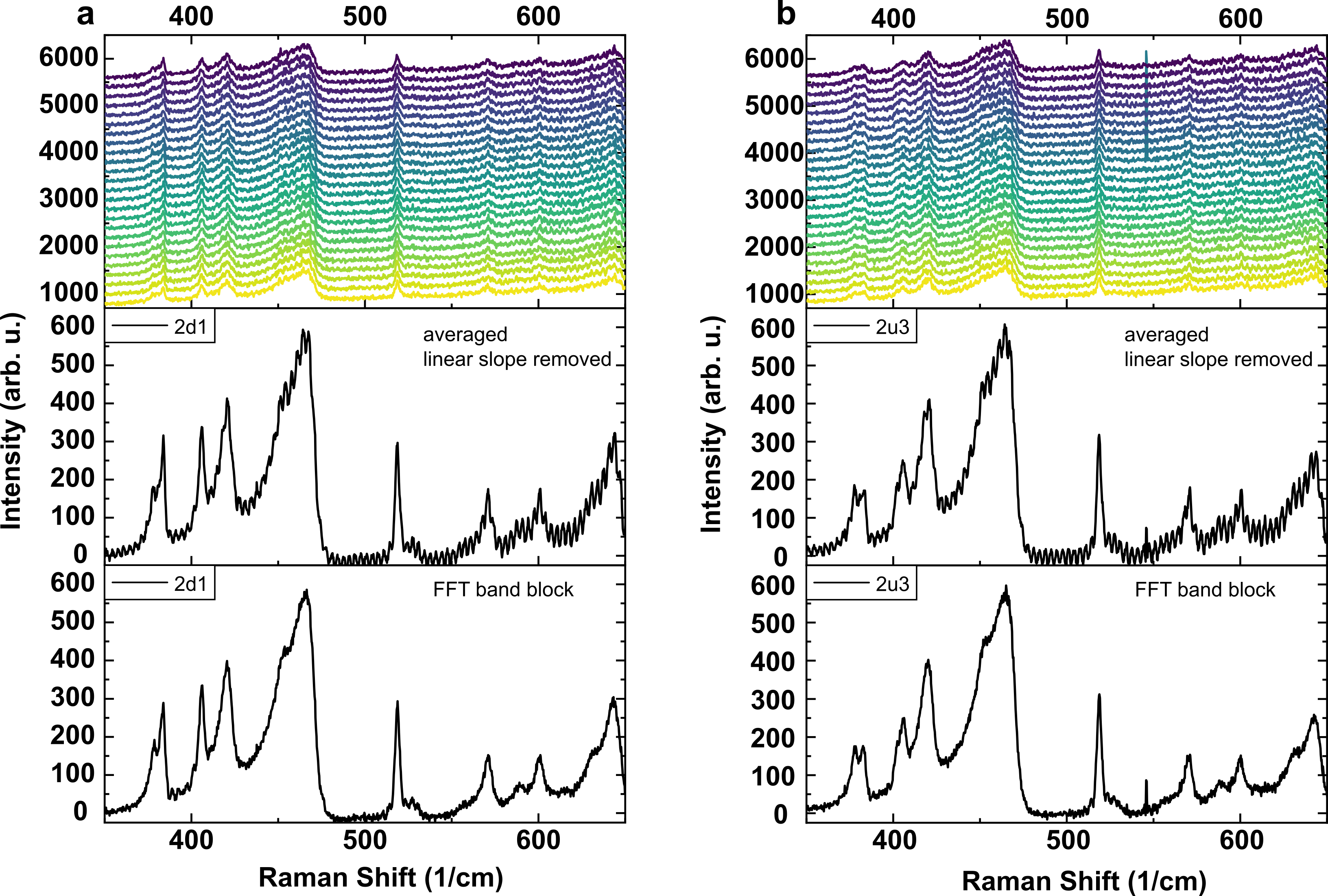}
	\caption{To process the near-rosonant Raman data and ``clean'' it, we first averaged over a set of 25 spectra, a 5 by 5 array of measurement points on the sample. Our measurement points are 0.33~µm appart, therefore, the spectrum is spatially averaged over close to 2~µm by 2~µm. A linear slope is removed form the spatially averaged spectrum to remove the PL background caused by the near-resonant excitation. The resulting spectrum (middle frame) shows shows a sine-like modulation over the whole spectrum caused by an etaloning effect in the setup. This is removed by using a narrow Fast-Fourier-Transformation band block. With this range, the etaloning is removed with little information loss in the data. This process is done for all spatially averaged near-resonant Raman data and shown as an example here for the up (a) and down (b) domains.}
	\label{fig:SI_DataProcessing}
\end{figure}

\end{document}